\documentclass[aip, amsmath,amssymb,preprint]{revtex4-1}

\usepackage{graphicx}
\usepackage{amsmath}
\usepackage[utf8]{inputenc}
\usepackage[T1]{fontenc}
\usepackage{mathptmx}
\usepackage{bm}

\newcommand{\wt}{\tilde{W}}
\newcommand{\bsigma}{\bm{\sigma}}
\newcommand{\bpi}{\bm{\pi}}

\begin{document}

\title{Prediction of native contacts in proteins from an out--of--equilibrium coevolutionary process}

\author{D. Oriani}
\affiliation{Department of Physics, Universit\`a degli Studi di Milano, via Celoria 16, 20133 Milano, Italy}
\author{M. Cagiada}
\altaffiliation[Present address: ]{Biology Institute, University of Copenhagen, Ole Maal\o es Vej 5, 2200 K\o benhavn N, Denmark}
\affiliation{Department of Physics, Universit\`a degli Studi di Milano, via Celoria 16, 20133 Milano, Italy}
\author{G. Tiana}
 \email{guido.tiana@unimi.it}
\affiliation{Department of Physics and Center for Complexity and Biosystems, Universit\`a degli Studi di Milano and INFN, via Celoria 16, 20133 Milano, Italy}

\date{\today}

\begin{abstract}
The analysis of coevolution of residues in homologous proteins is a powerful tool to predict their native conformation. The standard framework in which coevolutionary analysis is usually worked out is that of equilibrium Potts models, assuming that proteins have evolved for enough time to reach thermodynamic equilibrium in sequence space. Here we propose a model to describe correlations in sequences based on an explicit description of the evolutionary kinetics of proteins. We show that this procedure improves the correct prediction of native contacts with respect to equilibrium--based models.
\end{abstract}

\maketitle

\section{Introduction}

The determination of the native structure of proteins from their sequences is the 'holy grail' of protein science. Several strategies were developed to reach this goal, ranging from those based on the minimization of some effective energy function \cite{Das:2008} to homology modelling\cite{Al-Lazikani2001}. A particularly promising technique is that based on the analysis of the coevolution of pairs of residues in alignments of homologous sequences \cite{Gobel1994}, exploiting the fact that pairs of residues which are in contact in the native structure of a protein tend to evolve in a correlated way, thus leaving a signal in the associated sequence alignment. The problem of extracting this coevolutionary information and turning it into spatial restraints from which one can obtain the protein structure can be cast into an inverse Potts model. Different techniques were developed to solve the inverse problem in a computationally--efficient way, including those based on a mean--field approach \cite{Morcos2011,Jana2014}, on pseudo-likelihood estimation \cite{Ekeberg2013,Ekeberg2014} and on Boltzmann--learning algorithms \cite{Sutto2015,Franco2019}.

This is anyway a technique that has not yet reached its maturity. Despite important progresses, coevolutionary analysis alone is not still able to lead to a reliable prediction of protein structures and in the best cases it can predict of the order of 10\% of native contacts without false positives \cite{Ekeberg2014,Franco2019}. In fact, coevolutionary tools are often combined with other complementary methods to obtain better results \cite{Kamisetty2013,AlQuraishi2019a}. A non--trivial observation is that the use of techniques  that are intrinsically more correct for the inversion of the Potts model, for example Boltzmann learning as compared to the mean field approximation, does not lead to a drastic improvement of the predictions \cite{Franco2019}, suggesting that it is not only the inversion problem to affect the results.

One should thus not only work to refine the inversion algorithm, but also to take care of other aspects of the coevolutionary process. For example, it is known that binding \cite{Gueudre2016} or active \cite{Chakrabarti2010a} sites on the surface of proteins participate to coevolution, and could affect the prediction of intra--domain contacts by inserting spurious effects. In order to model these effects, one has to go beyond the standard modeling of protein evolution in terms of a simple Potts model.

Moreover, a key assumption that is implicitly done when modeling coevolution is that the input alignment is a realization of a probability distribution of sequences under stationary conditions. Even if it was suggested that evolution had enough time to reach equilibrium\cite{Rost1997}, this is an hypothesis extremely difficult to prove. If this is not the case, the variational principles that are at the basis of standard prediction techniques \cite{Morcos2011,Ekeberg2013,Sutto2015} will be not valid, and one will be compelled to describe protein alignments by time--dependent probability distributions. As a matter of fact, the statistical properties of homologous sequences resemble those of a glassy system \cite{Franco2019}, which can hardly reach equilibrium.

In the present work, we worked out an out--of equilibrium model for protein coevolution and implemented an inversion strategy based on the explicit reconstruction of phylogenetic trees by a simple stochastic process. For a given alignment of homologous sequences, we calculated the residue-residue empirical correlation functions, as in standard coevolutionary approaches, and we reconstructed the common ancestor and its evolutionary age by standard phylogenetic tools \cite{Felsenstein1985}. We then simulated the evolution of protein sequences by a minimal evolutionary model controlled by a set of energy parameters, obtaining a collection of offsprings and the associated correlation functions (see Sect. \ref{sect:model}). In this way, we take into account explicitly both the time-scale of the evolutionary process, not assuming equilibrium, and the phylogenetic correlations between simulated sequences. 

An optimization algorithm was then applied to the interaction energies to make the calculated correlation functions as similar as possible to the empirical ones, minimizing their square differences. However, the energy parameters of the model are mapped onto the correlation function by a stochastic process that does not seem to have an analytical solution. As a consequence, we could not rely on an exact form of the gradient to implement the optimization of the energies. For this reason, we worked out an approximated form of the evolutionary process and of associated gradient, as discussed in the Sect. \ref{sect:gradient}.

In Sect. \ref{sect:results} we show that taking into account the finite--time evolution of protein sequences improves the prediction of native contacts in the empirical alignment with respect to methods based on the equilibrium hypothesis.

\section{The evolutionary model} \label{sect:model}

While in standard coevolutionary analysis one inverts an equilibrium model, looking for the energy parameters for which the equilibrium properties of model sequences are as similar as possible to the experimental ones, here we inverted a time--dependent evolutionary model.

Given an ancestor sequence $\bsigma^0\equiv\{\sigma_i^0\}$, we want to simulate its evolution for a time $t$ as a stochastic branching process. In this way a tree of sequences is generated, and the sequences $\bsigma^{t}\equiv\{\sigma_i^{t}\}$ belonging to the last generation are regarded as a model for modern, homologous sequences. From each sequence in the process, a number $b$ of tentative offsprings is generated, each residue of the offspring being mutated  with probability $\mu$. A mutation is a random substitution of residue (or a gap, described as the 21st type of amino acid) at a given site with another one (or with a gap). The offspring is then accepted according to the only requirement that it represents a thermodynamically--stable protein, otherwise it is discarded.

Since evaluating the stability of a protein explicitly in conformational space would be computationally too time--consuming, we relied on the observation that stable proteins are those that display a sizable  gap between the energy of the native conformation and that of competing, unfolded conformations \cite{Shakhnovich1993,Shakhnovich1993a}. Since the energy of competing conformations does not depend on the specific sequence of the protein, its stability is uniquely defined by the energy $E_N$ of its native conformation. Thermodynamically--stable sequences are then those with $E_N\ll E_c$, where $E_c$ is the (self--averaging) lowest energy of denatured conformation.

As usually done in coevolutionary studies \cite{Morcos2011}, let's assume that the energy of a sequence in its native conformation is 
\begin{equation}
E_N(\sigma)=\sum_i h_i(\sigma_i)+\sum_{i<j} J_{ij}(\sigma_i,\sigma_j).
\end{equation}
The acceptance rate is defined as a sigmoidal function
\begin{equation}
w_{acc}(\bsigma\to\bsigma')=\theta(E_c-E_N(\bsigma'))\equiv\frac{1}{1+e^{-\kappa[E_c-E_N(\bsigma')]}},
\label{eq:acc}
\end{equation}
which depends on the energy parameters $h_i(\sigma)$ and $J_{ij}(\sigma,\pi)$ of the model. This model is a particularly simple implementation of the neutral theory of evolution \cite{Kimura1968}, in which  proteins are selectively neutral provided that they are thermodynamically stable. The jumping rate between sequences satisfies the principle of detailed balance, and then the distribution of sequence probability converges to equilibrium at large $t$.

Operatively, given the family of homologs, we reconstructed their common ancestor with the Phylip maximum--likelihood algorithm \cite{Felsenstein1985}. The evolutionary time $t$ is measured as mean number of mutations from the ancestor, according to the molecular--clock hypothesis \cite{Kimura1968}. The threshold energy $E_c$ is defined as the energy $E_N(\sigma^0)$ of the ancestor sequence, assuming that it is marginally stable. Given a set of energy parameters $h_i$ and $J_{ij}$, the evolutionary branching process is simulated for a time $t$. To avoid that the evolutionary tree becomes too wide, and thus difficult to store in a computer, we also introduce after each branching event, a number $a$ of replications with a single offspring ($b=1$).

\section{The optimization problem} \label{sect:gradient}

For any set of energy parameters $h_i(\sigma)$ and $J_{ij}(\sigma,\pi)$ we repeated $n_{sim}=??$ independent realizations of the evolutionary tree, calculating from all of them the one--point frequency $f_i(\sigma)=(\sum_n^{n_{sim}}M_n)^{-1}\sum_m \delta_{\sigma,\sigma^m_i}$ and the two--point correlation function $f_{ij}(\sigma,\pi)=(\sum_n^{n_{sim}}M_n)^{-1}\sum_{nm} \delta_{\sigma,\sigma_i^n}\delta_{\pi,\sigma_i^m}$, where $M_n$ is the number of sequences in the last generation of the simulation. 

Given the experimental quantities $f_{i}^{exp}(\sigma)$ and $f_{ij}^{exp}(\sigma,\pi)$ associated with the protein family, we looked for a model compatible with them by maximizing the quantity
\begin{eqnarray}
s&=&\sum_{i\sigma} \left[f_i(\sigma)-f_i^{exp}(\sigma)\right]^2 + \label{eq:s}\\
&+&\sum_{ij\sigma\pi}\left[f_{ij}(\sigma,\pi)-f_{ij}^{exp}(\sigma,\pi)\right]^2+ \lambda \sum_{ij\sigma\pi} J_{ij}(\sigma,\pi)^2,\nonumber
\end{eqnarray}
where the last term is a $L_2$ regularizer on the energy parameters, analogous to that used in equilibrium models \cite{Ekeberg2014}. The maximization is implemented by a steepest--descent algorithm such that at each step $h_i(\sigma) \to h_i(\sigma)-\eta ds / dh_i(\sigma)$ and $J_{ij}(\sigma,\pi) \to J_{ij}(\sigma,\pi)-\eta ds / dJ_{ij}(\sigma,\pi)$, where $\eta$ is a 'learning rate' that decreases 
with the number of steps $n$ as $\eta=\alpha/(1+n/n_0)^\beta$, with $\alpha=0.01$, $n_0=4000$ and $\beta=1$.

The non--trivial part of the algorithm described above is the calculation of the gradients $ds / dh_i(\sigma)$ and $ds / dJ_{ij}(\sigma,\pi)$. To obtain them, one can describe the evolutionary process as a Markov process where the probability $p(\bsigma,t)$ of a given sequence $\bsigma$ controlled by the master equation
\begin{equation}
\frac{dp(\bsigma,t)}{dt}=\sum_{\bsigma'} \left[ p(\bsigma',t) w(\bsigma'\to\bsigma) -  p(\bsigma,t) w(\bsigma\to\bsigma')\right],
\label{eq:master}
\end{equation} 
where 
the jumping rate can be written as
\begin{equation}
w(\bsigma'\to\bsigma)=\pi(\bsigma'|\bsigma) \cdot \frac{1}{1+e^{-\kappa[E_c-E_N(\bsigma')]}},
\label{eq:defw}
\end{equation}
where $\pi(\sigma'|\sigma)$ is the probability of generating a sequence $\bsigma'$ with a given Hamming distance (i.e., normalized number of mutations) from sequence $\bsigma$, and the fraction is the acceptance rate $w_{acc}$ of Eq. (\ref{eq:acc}). The solution of Eq. (\ref{eq:master}) is
\begin{equation}
p(\bsigma,t)=e^{\wt t} p(\bsigma,0),
\label{eq:solp}
\end{equation}
where $\wt(\bsigma,\bsigma')=w(\bsigma'\to\bsigma)-\delta_{\bsigma,\bsigma'}\sum_{\bpi} w(\bpi\to\bsigma)$ and $p(\bsigma,0)=\delta(\bsigma-\bsigma^0)$.

To calculate the gradient of $s$ defined by Eq. (\ref{eq:s}) one needs
\begin{equation}
\frac{\partial f_{kl}(\rho,\upsilon)}{\partial J_{ij}(\sigma,\pi)}  =   \frac{\partial}{\partial J_{ij}(\sigma,\pi)}  \sum_{\bm{\sigma}'} \delta_{\sigma'_k,\rho} \delta_{\sigma'_l,\upsilon}  p(\bm{\sigma}',t),
\end{equation}
in which the derivative acts only on the probability; using Eq. (\ref{eq:solp}) this derivative is
\begin{equation}
 \frac{\partial p(\bsigma,t)}{\partial J_{ij}(\sigma,\pi)} = t \sum_{\bsigma',\bsigma''}\left(e^{\wt t}\right)_{\bsigma\bsigma'} \left(\frac{\partial \wt}{\partial J_{ij}(\sigma,\pi)}\right)_{\bsigma'\bsigma''}  p(\bsigma'',0)
\end{equation}
Using the definition of $\wt$,  Eq. (\ref{eq:defw}) and the relation $ \theta'(E_c-E_N(\bsigma^0)) =\kappa/4$ one obtains
\begin{equation}
\frac{\partial f_{kl}(\rho,\upsilon)}{\partial J_{ij}(\sigma,\pi)}  =  G_1+G_2
\end{equation}
with
\begin{eqnarray}
G_1 &=&  \frac{t\kappa}{4}\delta_{\sigma^0_i,\sigma} \delta_{\sigma^0_j,\pi} \sum_{\bsigma',\bsigma'',\bsigma'''} \delta_{\sigma'_k,\rho}\delta_{\sigma'_l,\upsilon} \times \nonumber\\
&\times&\left(e^{\wt t}\right)_{\bsigma'\bsigma''}  \cdot \pi(\bsigma'''|\bsigma'')p(\bsigma''',0)\nonumber\\
G_2&=&t \left[ \sum_{\bsigma'\bsigma''} \delta_{\sigma'_k,\rho}\delta_{\sigma'_l,\upsilon} \left(e^{\wt t}\right)_{\bsigma'\bsigma''} p(\bsigma'',0)\right]\times \nonumber\\
&\times&  \left[ \sum_{\bsigma'\bsigma''}  \theta'(E_c-E_N(\bsigma')) \delta_{\sigma'_i,\sigma}\delta_{\sigma'_j,\pi} \right. \times\nonumber\\
&\times&\left. \pi(\bsigma'|\bsigma'') p(\bsigma'',0) \right].
\label{eq:g1g2}
\end{eqnarray}
 
We need to write $G_1$ and $G_2$ in a form that can be calculated numerically from the simulated sequences. The quantity $G_2$ is easier to calculate. It is composed of two factors; the former contains the propagator (\ref{eq:solp}) applied to the probability distribution associate with the ancestor, that gives $p(\bsigma',t)$. Thus, this term gives the average
\begin{equation}
\langle \delta_{\rho,\sigma^{t}_k}\delta_{\upsilon,\sigma^{t}_l} \rangle\equiv \sum_{\bsigma'} \delta_{\sigma'_k,\rho}\delta_{\sigma'_l,\upsilon} p(\bsigma',t)
\end{equation}
over sequences $\bsigma^t$ generated at time $t$ of the simulation. The latter term of $G_2$ can be written as
\begin{equation}
 \sum_{\bsigma'} \pi(\bsigma'|\bsigma^0) \theta'(E_c-E_N(\bsigma')) \delta_{\sigma'_i,\sigma}\delta_{\sigma'_j,\pi}, 
\end{equation}
where the probability $\pi(\bsigma' | \bsigma)$ that sequence $\bsigma'$ displays a Hamming distance $d(\bsigma')$ from sequence $\bsigma^0$ can be regarded as a binomial distribution ${{N}\choose{d}} \mu^d(1-\mu)^{N-d}$. The most important contribution to this term comes from sequences with energy close to $E_c$ (within $\kappa$ from it, because $\theta'$ goes to zero beyond it) and with Hamming distance $N\mu$ from the ancestor (because here the binomial distribution has its maximum), corresponding to sequences $\sigma^{t}$ at the bottom of the tree. Thus, 
\begin{equation}
G_2\approx\langle \delta_{\rho,\sigma^{t}_k}\delta_{\upsilon,\sigma^{t}_l} \rangle\cdot \langle  \theta'(E_c-E_N(\bsigma')) \delta_{\sigma^{t}_i,\sigma}\delta_{\sigma^{t}_j,\pi} \rangle.
\end{equation}

\begin{figure}[h]
  \includegraphics[width=\columnwidth]{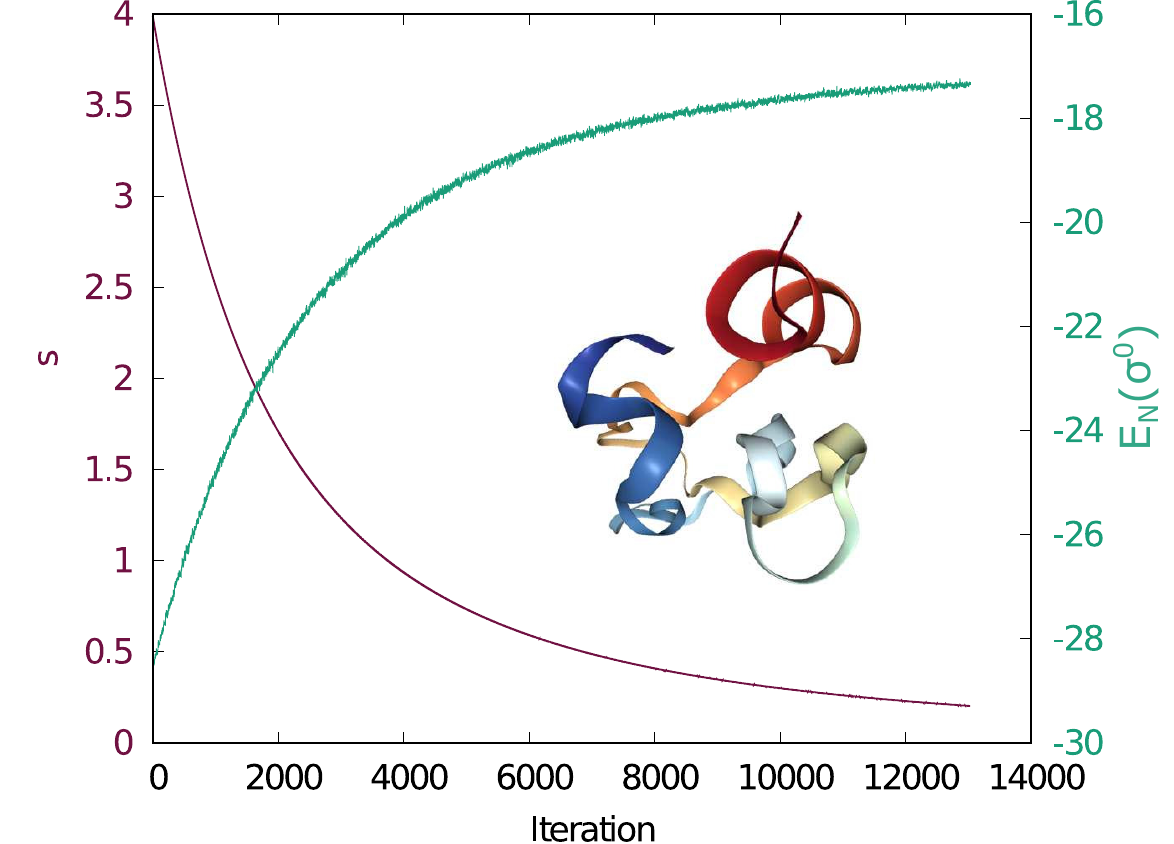}
  \caption{An example of minimization of the loss function $s$ for protein 1BPI (left axis) and the native energy $E_N(\sigma^0)$ of the ancestor protein (right axis) as a function of the number of iterations of the optimization algorithm.}
  \label{fig:loss_fun_plot}
\end{figure}

The term $G_1$ is more cumbersome, but acts only a minority of elements of the energy tensor, namely those that in a given pair of sites have the same residues as the ancestor, because of the Kronecker's delta in the first line of Eq. (\ref{eq:g1g2}). Making the (strong) assumption that the mutability is small, and then $\pi(\bsigma'''|\bsigma'')\approx\delta(\bsigma''-\bsigma''')$, one obtains
\begin{equation}
G_1 \approx  \frac{t\kappa}{4}\delta_{\sigma^0_i,\sigma} \delta_{\sigma^0_j,\pi} \langle \delta_{\rho,\sigma^{t}_k}\delta_{\upsilon,\sigma^{t}_l} \rangle,
\end{equation}
which would be equal to $-G_1$ if the latter average of $G_2$ were restricted to $\bsigma^0$. Then, defining $\langle\cdot\rangle'$ as the average over all offspring sequences except those displaying residue $\sigma_i^0$ at position $i$ and $\sigma_j^0$ at position $j$, one has
\begin{equation}
\frac{\partial f_{kl}(\rho,\upsilon)}{\partial J_{ij}(\sigma,\pi)}  \approx \langle \delta_{\rho,\sigma^{t}_k}\delta_{\upsilon,\sigma^{t}_l} \rangle\cdot \langle  \theta'(E_c-E_N(\bsigma')) \delta_{\sigma^{t}_i,\sigma}\delta_{\sigma^{t}_j,\pi} \rangle'.
\end{equation}
Using this approximation for the gradient, the quantity $s$ one has to minimize decreases
quite efficiently, as shown in Fig. \ref{fig:loss_fun_plot}.

\section{Prediction of native contacts} \label{sect:results}

To evaluate the performance of the present non-equilibrium method with respect to standard equilibrium approaches, and specifically pseudo--likelihood inversion, we applied them to four different proteins, namely bovine pancreatic trypsin inhibitor (pdb code 1BPI), immunophilin immunosuppressant (pdb code 1FKJ), acyl-coenzyme A binding protein (pdb code 2ABD) and an aminopeptidase (pdb code 5IB9). We used the mutability parameter $\mu=0.15$, a sigmoid steepness $\kappa=5$ and a regolarizer parameter $\lambda=0.01$. The branching ratio was chosen in the range between $b=30$ and $b=120$, in order to keep the number of sequences generated at the bottom of the phylogenetic tree between $5000$ to $90000$. This number is quite variable for fixed parameters depending on the specific protein, and thus the value of $b$ should be tuned too avoid that it becomes too small, resulting in a bad statistics, or too large, causing computational storage problems.

All the simulations were performed using as starting
potential that obtained by the maximization of the pseudolikelihood, against which we
compare our final potentials in order to evaluate if the non-equilibrium method can be
used to refine the predictions of this equilibrium method.

For each protein we could reach convergence for the loss function $s$ in $\sim 10^4$ iterations. After the minimization, the energy parameters $h$ and $J$ changed consistently, as shown for example from the fact that the native energy $E_N(\sigma_0)$ of the ancestor (measured in the constant units of $\lambda^{-2}$), changed largely along the optimization (cf. Fig. 1 for the case of 1BPI).

After the optimization, each contact $i-j$ is scored with the direct information\cite{Morcos2011} $DI_{ij}=\sum_{\sigma\pi}p^{dir}_{ij}(\sigma,\pi)\log[ p^{dir}_{ij}(\sigma,\pi) / f_i(\sigma)f_i(\pi)]$, where $p^{dir}_{ij}(\sigma,\pi)=Z_{ij}^{-1}\exp[-h_i(\sigma)-h_j(\pi)-J_{ij}(\sigma,\pi)]$, normalized to its larges value $DI_{max}$, and ordered following the value of their $DI/DI_{max}$. The rationale is that contacts with largest direct information are native, that is are present in the crystallographic structure (true positive predictions) while those with lowest direct information are not (true negatives).

\begin{figure}[h]
  \includegraphics[width=\columnwidth]{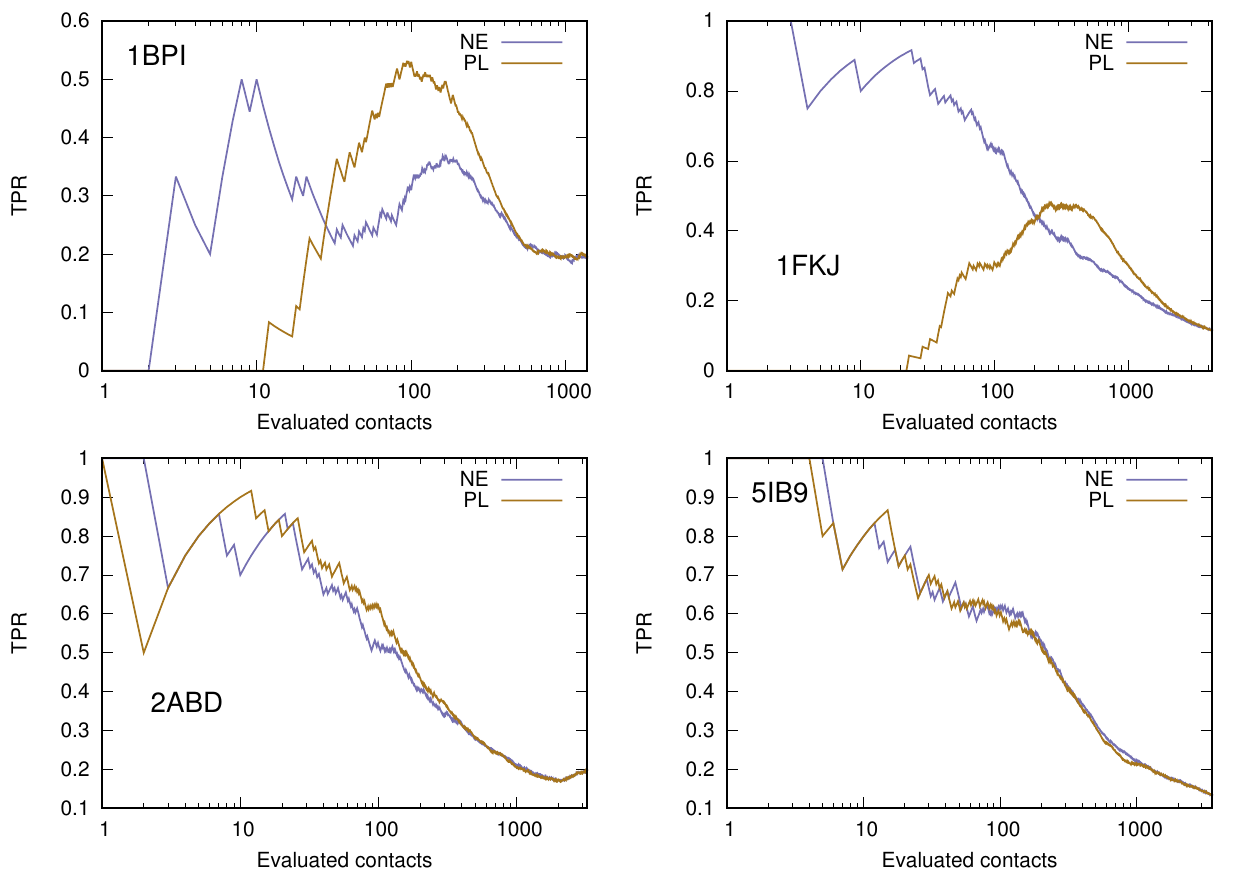}
  \caption{Comparison of true positive ratio (TPR) for four different proteins. The blue lines
  perindicate the results of our non-equilibrium method (NE) while the yellow ones denote the
  results of the pseudolikelihood maximization method (PL).}
  \label{fig:tprpanel}
\end{figure}

In Fig. \ref{fig:tprpanel} it is shown that true positive ratio (TPR), defined as the fraction of true positive contacts among those whose relative DI is larger than a given threshold over all contacts whose relative DI is above that threshold, as a function of the number of contacts. In an ideal prediction the TPR assumes the value 1 up to the number of true native contacts, and then drops to zero. The figure shows the results for both our non--equilibrium method (NE) and for the equilibrium pseuso--likelihood (PE). For proteins 1BPI and 1FKJ the prediction of the equilibrium method is quite poor, and the non--equilibrium corrections improves it considerably. For 2ABD and 5IB9, the prediction of the equilibrium methods is rather good, and the non-equilibrium corrections improves negligibly this result. Using other estimators, as the Frobenius norm instead of the direct information, or other equilibrium algorithms, like Boltzmann learning, does not affect substantially this results (data not shown, cf. also ref. \cite{Franco2019}).

\begin{figure}[h]
  \includegraphics[width=\columnwidth]{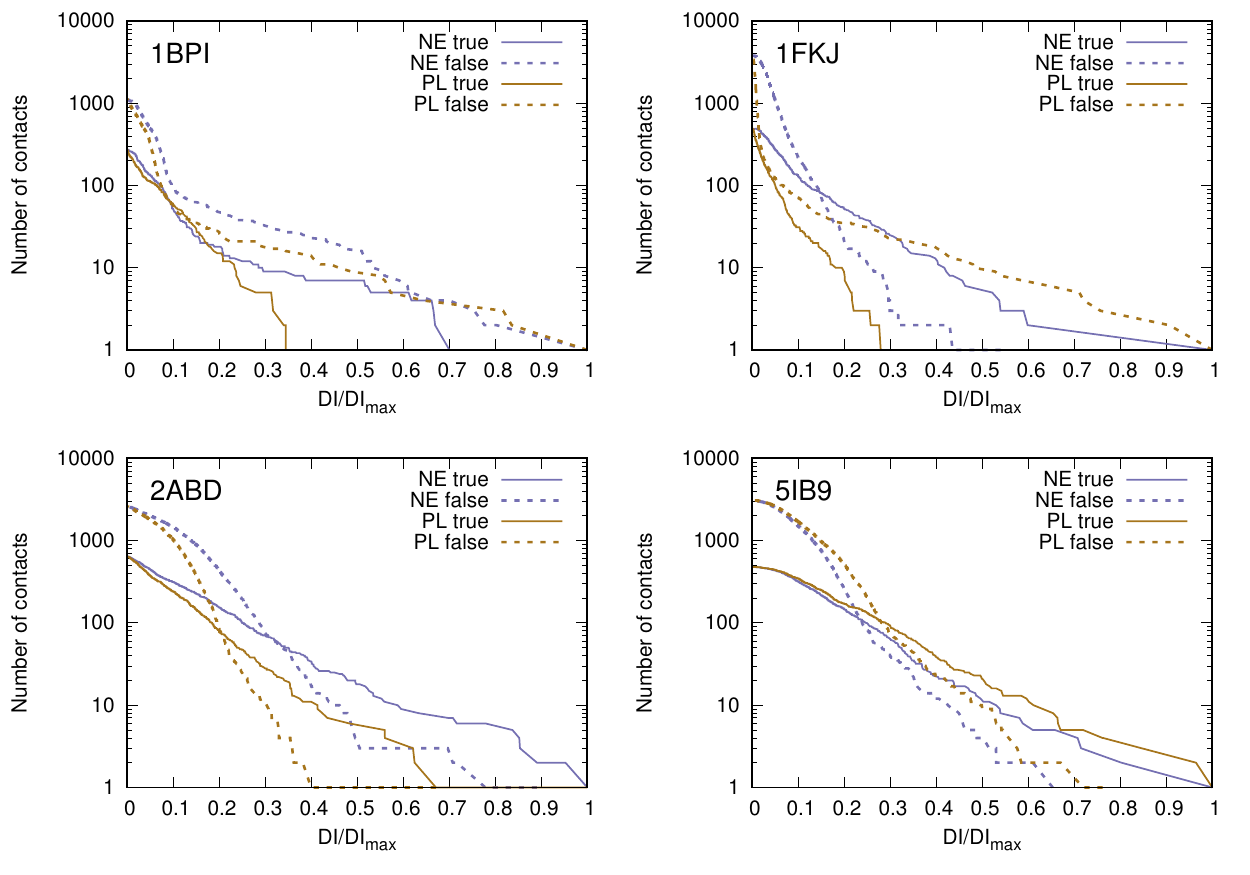}
  \caption{Comparison of predicted contacts for four different proteins. The blue lines
  pertain the results of our non-equilibrium method (NE) while the yellow ones denote the
  results of the pseudolikelihood maximization method (PL). Solid lines indicate true positive
  contacts, while dashed lines show the false positive contacts.}
  \label{fig:threshpanel}
\end{figure}

Figure  \ref{fig:threshpanel}  displays the number of true and false positives for NE and PL as a function of the threshold of relative direct information, For 1BPI, 1FKJ and 2ABD the effect of the non--equilibrium method is that of increasing the number of true positives at all values of $DI/DI_{max}$, while for 5IB9 this remains similar or decreases slightly. For 1FKJ one can also observe a consistent reduction of the number of false positives using the out--of--equilibrium algorithm, although this does not seem to be a general result.

\begin{figure}[h]
  \includegraphics[width=\columnwidth]{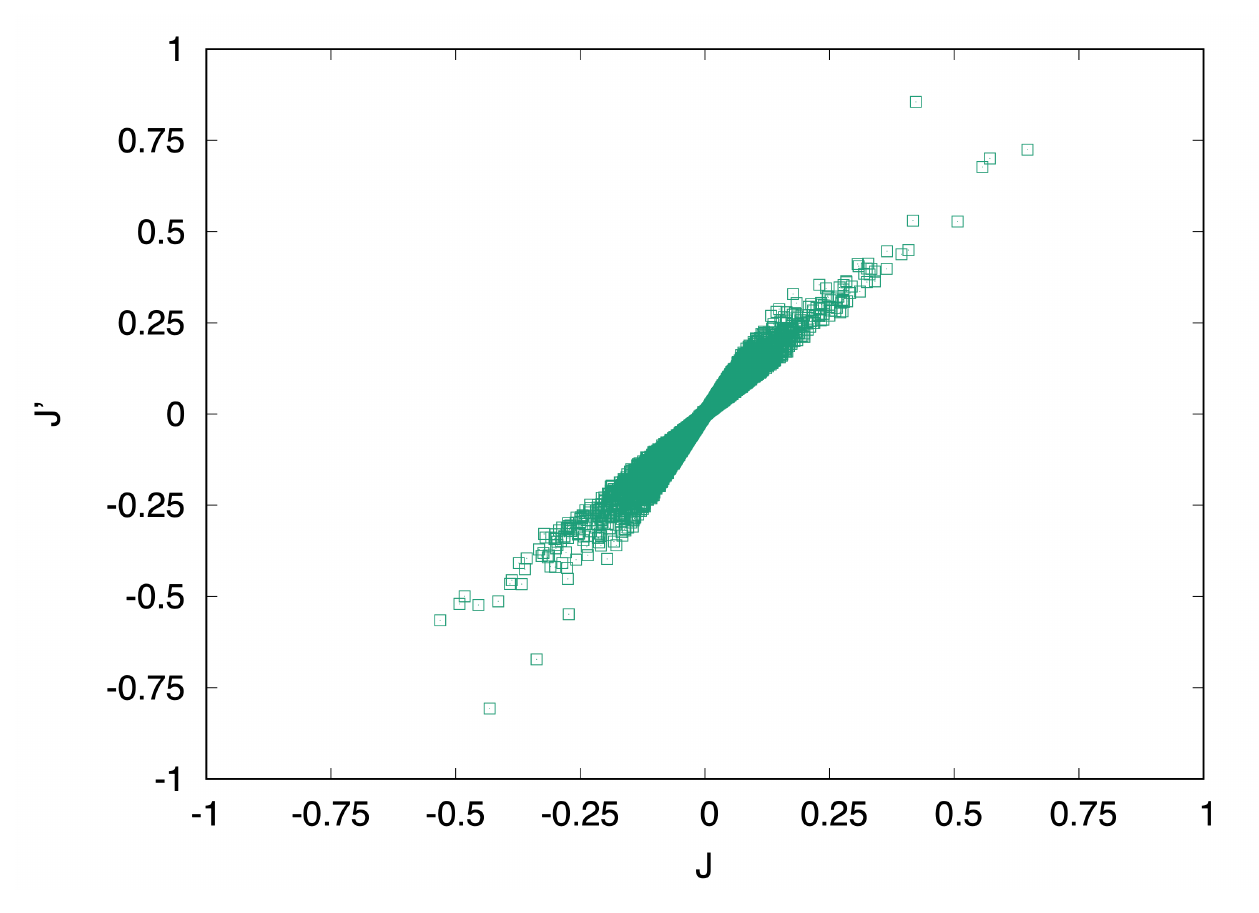}
  \caption{Comparison of the final two-body coevolutionary potentials obtained for two
  separate simulations on 1BPI. The simulations were carried out with the same parameters
  and conditions.}
  \label{fig:tensorcomp}
\end{figure}

An interesting question about this method is that regarding the uniqueness of the
obtained coevolutionary potentials. While in the equilibrium case this is guaranteed by the convexity of the function to be minimized, this is not the case here. Thus,  we investigated this issue numerically,  repeating two indipendent optimizations for
protein 1BPI and we compared the two-body potentials $J_{ij}(\alpha,\beta)$, obtaining the value 0.9729 for the
Pearson correlation coefficient (cf. Fig. \ref{fig:tensorcomp}). 

\section{Conclusions}

We proposed a scheme to infer spatial contacts in proteins based on coevolutionary data that does not rely on the assumption that sequences are in equilibrium is sequence space. This methods improves the prediction when standard equilibrium strategies fail and do not worsen the prediction when equilibrium ones are already good.

This is indeed a scheme that must be regarded as a correction to equilibrium methods, because it requires a reasonable starting guess of the energy parameters. If one starts from random (or zero) initial energy parameters, it is very unlikely that the simulated evolution can produce offsprings with inter--residue correlations comparable with the experimental ones. In this case typically the iterative optimization gets lost in the huge parameter space.

One should only be aware that the out--of--equilibrium strategy is much more computational requiring than the usual equilibrium ones, like the mean--field and the pseudo--likelihood optimization. On a desktop computer, using 8 cores in parallel, the runtime of a calculation varied between $3$ and $7$ days.  The ideal application of it is then to refine the cases when equilibrium methods fail.

\end{document}